# Adaptive Restructuring of Merkle and Verkle Trees for Enhanced Blockchain Scalability

Oleksandr Kuznetsov [1,2*], Dzianis Kanonik [1], Alex Rusnak [1], Anton Yezhov [1], Oleksandr Domin[1]

[1] Proxima Labs, 1501 Larkin Street, suite 300, San Francisco, USA
[2] Department of Political Sciences, Communication and International Relations, University of Macerata, Via Crescimbeni, 30/32, 62100 Macerata, Italy

*Corresponding author. E-mail(s): kuznetsov@karazin.ua

Contributing authors: alex@proxima.one

**Abstract:** The scalability of blockchain technology remains a pivotal challenge, impeding its widespread adoption across various sectors. This study introduces an innovative approach to address this challenge by proposing the adaptive restructuring of Merkle and Verkle trees, fundamental components of blockchain architecture responsible for ensuring data integrity and facilitating efficient verification processes. Unlike traditional static tree structures, our adaptive model dynamically adjusts the configuration of these trees based on usage patterns, significantly reducing the average path length required for verification and, consequently, the computational overhead associated with these processes. Through a comprehensive conceptual framework, we delineate the methodology for adaptive restructuring, encompassing both binary and non-binary tree configurations. This framework is validated through a series of detailed examples, demonstrating the practical feasibility and the efficiency gains achievable with our approach. Moreover, we present a comparative analysis with existing scalability solutions, highlighting the unique advantages of adaptive restructuring in terms of simplicity, security, and efficiency enhancement without introducing additional complexities or dependencies. This study's implications extend beyond theoretical advancements, offering a scalable, secure, and efficient method for blockchain data verification that could facilitate broader adoption of blockchain technology in finance, supply chain management, and beyond. As the blockchain ecosystem continues to evolve, the principles and methodologies outlined herein are poised to contribute significantly to its growth and maturity.

**Keywords:** Blockchain Scalability, Merkle and Verkle Trees, Adaptive Restructuring, Data Verification, Efficiency Optimization, Blockchain Architecture

## 1. Introduction

The advent of blockchain technology has heralded a new era in digital transactions, offering unparalleled security, transparency, and decentralization [1]. At its core, blockchain leverages cryptographic principles to create a distributed ledger system, where data integrity and transaction veracity are maintained across a network of nodes without the need for a central authority [2]. This innovative approach has found applications far beyond its initial cryptocurrency origins, extending into finance, supply chain management, healthcare, and more [3].

However, as blockchain technology ventures into more complex and demanding applications, it encounters a fundamental challenge that threatens its broader adoption: scalability [4]. The scalability issue primarily revolves around the capacity of a blockchain network to handle a large volume of transactions quickly and efficiently. Current blockchain architectures, while robust and secure, are hampered by their inherent design, which leads to bottlenecks in transaction processing and data verification [5,6]. These limitations not only increase transaction costs but also extend the time required to achieve consensus across the network, thereby reducing the system's overall throughput.

**1.1. The Blockchain Paradigm and the Challenge of Scalability**

In the burgeoning landscape of blockchain technology, Ethereum stands out as a beacon of innovation and application diversity. The network's capacity to support a wide array of decentralized applications (dApps) and smart contracts has positioned it at the forefront of blockchain development. This prominence is underscored by the data depicted in Figure 1, which reveals a significant uptick in the total number of unique Ethereum addresses over the past year, surging from 219.26 million to 254.66 million [7].

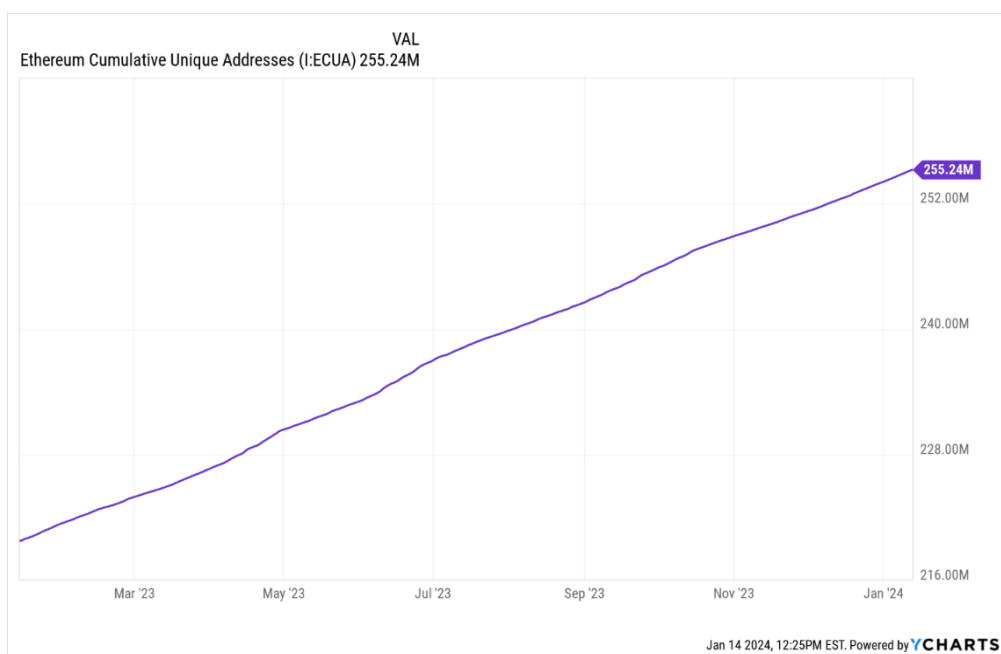

**Figure 1:** Growth in Total Unique Ethereum Addresses

However, the rapid expansion of the Ethereum network brings to light the pressing challenges associated with managing an ever-growing blockchain ecosystem. The core issue revolves around the efficient verification and management of data within the blockchain's infrastructure, a task that becomes increasingly complex as the network scales. The juxtaposition of the network's growth against the slight decrease in daily active Ethereum addresses, as shown in Figure 2, further complicates this challenge [8]. Despite a year-over-year decrease of 1.70% in daily active addresses, the fact remains that a mere 0.17% of all unique addresses engage with the network on a daily basis. This discrepancy between the total number of addresses and the proportion of active participants underscores a critical aspect of blockchain management: the network's activity is highly concentrated within a relatively small segment of the overall user base.

This concentration of activity presents a unique set of challenges in optimizing the blockchain's state tree. The state tree must be updated frequently to reflect the transactions and interactions occurring within the network, a process that is predominantly influenced by the small fraction of actively participating addresses. The need to efficiently manage and verify these updates, without compromising the integrity or performance of the network, is paramount.

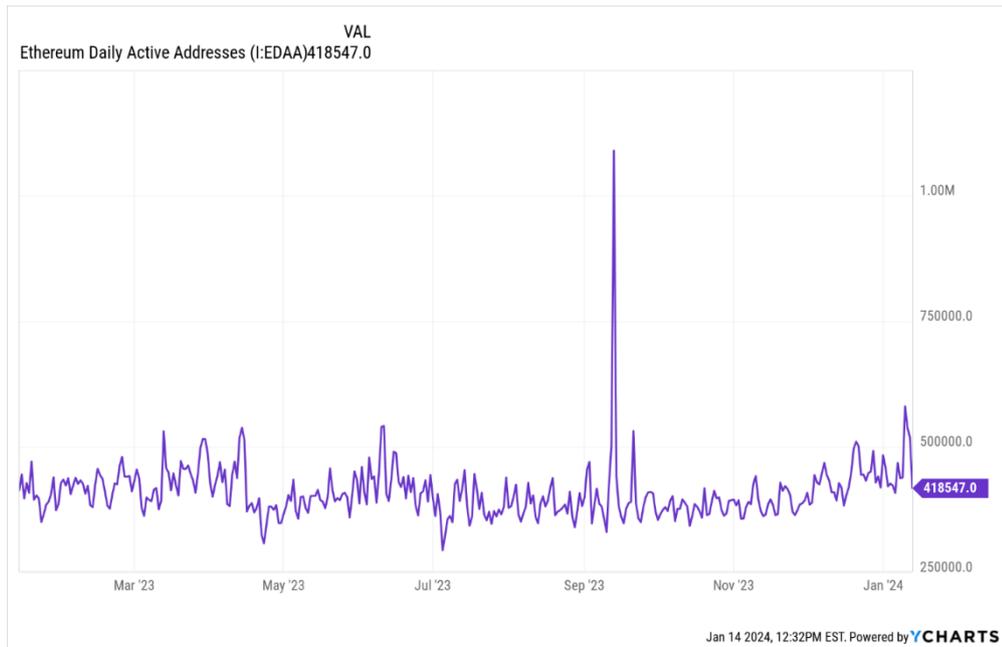

**Figure 2:** Daily Active Ethereum Addresses

Given this backdrop, our research aims to address the pressing need for an optimized approach to managing the blockchain's state tree, particularly within the context of Ethereum's rapidly expanding network. The goal of our study is to explore innovative methods for restructuring Merkle and Verkle Trees adaptively, thereby enhancing the efficiency of data verification processes. By focusing on dynamic adjustments to tree configurations in response to usage patterns, we seek to minimize verification path lengths and reduce the computational overhead associated with maintaining data integrity. This research endeavor not only aims to bolster the scalability of blockchain systems but also to contribute to the ongoing discourse on optimizing blockchain infrastructure for the next generation of decentralized applications.

**1.2. State of the art**

In the realm of blockchain scalability, Kottursamy et al. (2023) [9] introduce a novel blockchain architecture termed Mutable Block with Immutable Transaction (MBIT), aiming to enhance scalability through a trapdoor cryptographic hash function for quantifying unspent coins. While their approach significantly reduces verification and confirmation times, it primarily focuses on transaction efficiency without addressing the broader scalability challenges related to blockchain's data structure and state management.

Li et al. (2023) [10] propose PRI, a Payment Channel Hubs (PCH) solution enhancing privacy, reusability, and interoperability for blockchain scalability. Despite its innovative approach to solving the deposit lock-in problem and supporting multi-party participation, PRI's reliance on trusted hardware and its limited scope in addressing the fundamental architectural scalability issues of blockchain remain unaddressed.

Nasir et al. (2022) [11] provide a systematic review of scalable blockchains, identifying key strategies for enhancing blockchain capabilities and analyzing scalability solutions. Their work highlights the multifaceted nature of blockchain scalability but leaves a gap in practical implementation strategies for optimizing blockchain data structures, particularly in the context of state management and verification processes.

Sanka and Cheung (2021) [12] offer a comprehensive review of blockchain scalability issues and solutions, emphasizing the need for efficient consensus mechanisms and system throughput improvements. While their analysis sheds light on the scalability challenges, the exploration of adaptive data structures for optimizing blockchain's underlying architecture is not thoroughly explored.

Sharma et al. (2023) [13] introduce BLAST-IoT, a blockchain-assisted scalable trust model for the Internet of Things (IoT), focusing on secure dissemination and storage of trust information. Their model addresses scalability in the context of IoT devices but does not extend to the broader blockchain scalability challenges, particularly in relation to adaptive restructuring of blockchain data structures.

Wang and Wu (2024) [14] present Lever-FS, a validation framework for intensive blockchain validation, achieving scalability through optimistic execution and dispute resolution. While their work advances the scalability of validation processes, it does not directly tackle the optimization of blockchain's state tree structure for overall network efficiency.

Wang et al. (2023) [15] propose a scalable, efficient, and secured consensus mechanism for Vehicle-to-Vehicle (V2V) energy trading, leveraging blockchain technology. Their consensus mechanism addresses scalability in the specific context of V2V energy trading but does not address the broader application of scalable data structures within the blockchain.

Xiao et al. (2024) [16] develop CE-PBFT, a high availability consensus algorithm for large-scale consortium blockchain, focusing on improving system throughput and reducing latency. While their algorithm enhances consensus efficiency, the exploration of adaptive and scalable blockchain data structures remains an area for further research.

Yu et al. (2023) [17] introduce OverShard, a full sharding approach for scaling blockchain, which significantly improves throughput and reduces confirmation latency. However, the application of sharding to optimize blockchain's state tree and the exploration of adaptive restructuring techniques are not fully addressed.

Zhen et al. (2024) [18] propose a dynamic state sharding blockchain architecture for scalable and secure crowdsourcing systems, addressing the scalability and security of blockchain in crowdsourcing applications. While their architecture offers improvements in throughput and security, the potential for adaptive restructuring of blockchain data structures to further enhance scalability is not explored.

Transitioning from the broader challenges of blockchain scalability, we delve into the specific realm of tree-based data structures within blockchain technology. These structures, notably Merkle and Verkle trees, are pivotal for ensuring data integrity, enhancing verification processes, and optimizing storage within blockchain systems. The following literature review highlights significant advancements and identifies gaps that our research aims to fill.

Ayyalasomayajula and Ramkumar (2023) [19] explore the optimization of Merkle Tree structures, comparing linear and subtree implementations. Their findings favor the subtree method for its efficiency in handling large-scale databases. However, their work primarily focuses on theoretical advantages without addressing the practical challenges of integrating these optimized structures into existing blockchain frameworks.

Jeon et al. (2023) [20] introduce a hardware-accelerated approach for generating reusable Merkle Trees for Bitcoin blockchain headers, significantly reducing execution time and power consumption. While their solution enhances the efficiency of block candidate generation, it is tailored to Bitcoin and does not explore the adaptability of their approach to other blockchain architectures or the potential for further optimization in tree structure.

Jing, Zheng, and Chen (2021) [21] provide a comprehensive review of Merkle Tree's technical principles and applications across various fields. Their work underscores the versatility and potential of Merkle Trees but stops short of proposing innovative methods for dynamic restructuring or optimization of these trees in response to the evolving needs of blockchain systems.

Knollmann and Scheideler (2022) [22] present a self-stabilizing protocol for the Hashed Patricia Trie, a distributed data structure enabling efficient prefix search. Their protocol addresses self-stabilization in distributed systems but does not explore the scalability implications of their data structure within the broader context of blockchain technology.

Lin and Chen (2023) [23] propose a file verification scheme based on Verkle Trees, highlighting the efficiency and security benefits over traditional Merkle Trees. While their work demonstrates the potential of Verkle Trees in file verification, the exploration of these trees for broader blockchain scalability and optimization remains limited.

Liu et al. (2021) [24] offer systematic insights into Merkle Trees, emphasizing their role in blockchain data verification and retrieval. Their discussion on the advantages and applications of Merkle Trees provides a solid foundation but lacks a detailed exploration of innovative approaches to enhance these trees' efficiency and scalability in blockchain systems.

Mardiansyah, Muis, and Sari (2023) [25] introduce the Multi-State Merkle Patricia Trie (MSMPT) for high-performance data structures in multi-query processing. Their work addresses performance and efficiency in lightweight blockchain but does not delve into the scalability challenges of more complex blockchain systems.

Mitra, Tauz, and Dolecek (2023) [26] propose the Graph Coded Merkle Tree to mitigate Data Availability Attacks in blockchain systems. While their approach offers a novel solution to a specific problem, the broader application of their design for general blockchain scalability and data structure optimization is not addressed.

Zhao et al. (2024) [27] focus on minimizing block incentive volatility through Verkle tree-based dynamic transaction storage. Their innovative approach addresses a crucial aspect of blockchain economics but does not explore the structural optimization of Verkle Trees for enhanced scalability and efficiency in blockchain systems.

Our research fills these gaps by proposing adaptive restructuring techniques for Merkle and Verkle Trees, aiming to enhance blockchain scalability, optimize data verification and storage processes, and provide a flexible framework adaptable to various blockchain architectures and applications.

### 1.3. Our contribution

This work introduces a novel approach to optimizing tree-based data structures within blockchain technology, focusing on adaptive restructuring techniques for Merkle and Verkle Trees. Our contributions are twofold: First, we propose a dynamic restructuring algorithm that enhances the scalability and efficiency of blockchain systems by optimizing the verification and storage processes. Second, we extend the applicability of these optimized tree structures beyond traditional blockchain applications, demonstrating their versatility in various blockchain architectures and scenarios. Through rigorous analysis and experimentation, our research addresses the critical scalability challenges faced by blockchain technology, offering a scalable, efficient, and adaptable solution.

**1.4. Article structure**

The structure of this article is designed to provide a comprehensive overview of our research and findings. Section 2 conceptualizes the problem of blockchain scalability and the role of tree-based data structures in addressing this challenge. Section 3 introduces our idea for optimizing trees in blockchain, detailing the theoretical foundation of our approach. Section 4 evaluates the efficiency of adaptive Merkle trees through analytical and empirical methods. Section 5 describes the algorithm for Merkle Tree restructuring, followed by Section 6, which presents examples of the algorithm's execution in various scenarios. Section 7 delves into the specifics of path encoding in adaptive Merkle Trees, and Section 8 explores the enhancement of Verkle Trees through adaptive restructuring. The discussion in Section 9 synthesizes our results, comparing them with existing solutions and highlighting our contribution to the field. Finally, the conclusion in Section 10 summarizes our research contributions and outlines future directions for this promising area of study.

**2. Conceptualizing the Problem**

The core issue addressed in this research is the optimization of tree structures in blockchain systems for efficient and cost-effective verification. Currently, blockchain data is stored in balanced trees, with Merkle paths for data verification being approximately equal in length and complexity across all data. This uniformity results in a consistent verification cost and complexity, regardless of the frequency of data use.

Figure 3 depicts a balanced Merkle Tree, a fundamental data structure used in blockchain for ensuring data integrity. Each leaf node (A-P) represents a block of data with a unique hash value, while the non-leaf nodes (AB, CD, etc.) are hashes of their respective child nodes. The root node (ABCDEFGHIJKLMNOP) encompasses the entire tree's hash, providing a single point of reference for the entire dataset's integrity.

The Merkle Tree's structure ensures that any alteration in a single data block can be quickly detected by recalculating the hashes up the tree to the root. However, this balanced structure, while efficient in evenly distributing the data, does not account for the frequency of data access or modification (frequency is indicated in brackets). As a result, frequently used data and rarely accessed data have the same level of complexity and cost in terms of verification, leading to inefficiencies in resource utilization.

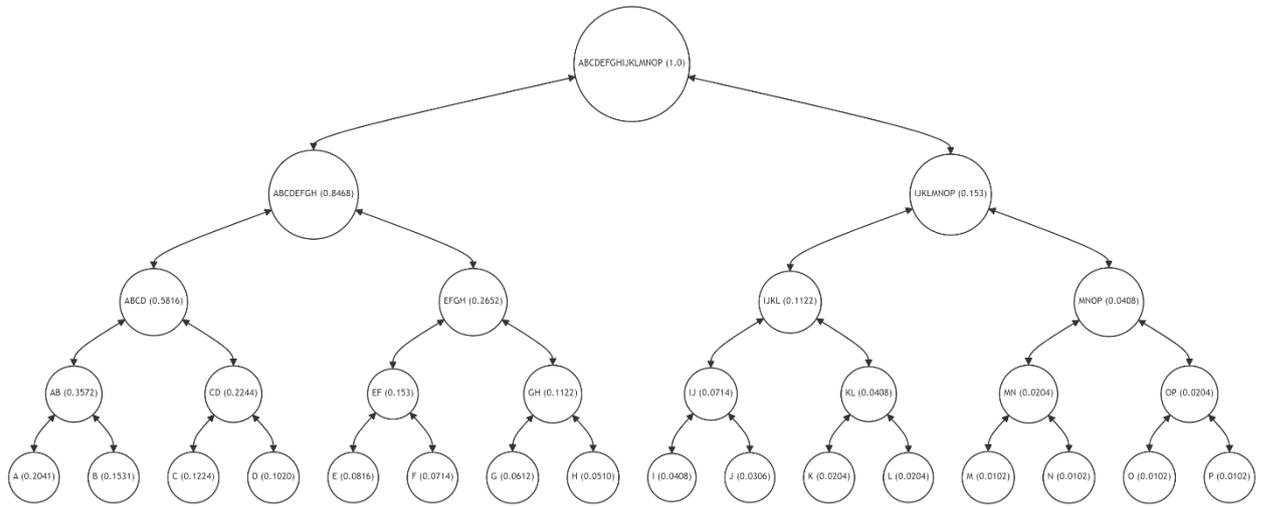

**Figure 3:** Balanced Merkle Tree Structure

Figure 4 highlights the Merkle Path (nodes B, CD, EFGH, IJKLMNOP) for verifying the integrity of leaf node A (with a high frequency of 0.2041). The Merkle Path is marked in red, indicating the nodes whose hashes are required to verify A's integrity up to the root. The leaf node A and the root are highlighted in green, while the intermediate nodes (AB, ABCD, ABCDEFGH) involved in hash calculations are in yellow.

The verification process involves recalculating and comparing the hashes from node A up to the root, ensuring data integrity. However, this method, while straightforward, applies the same verification complexity to all data, regardless of usage frequency. This "one-size-fits-all" approach is suboptimal, especially for data that is accessed and modified frequently, as it incurs unnecessary computational overhead.

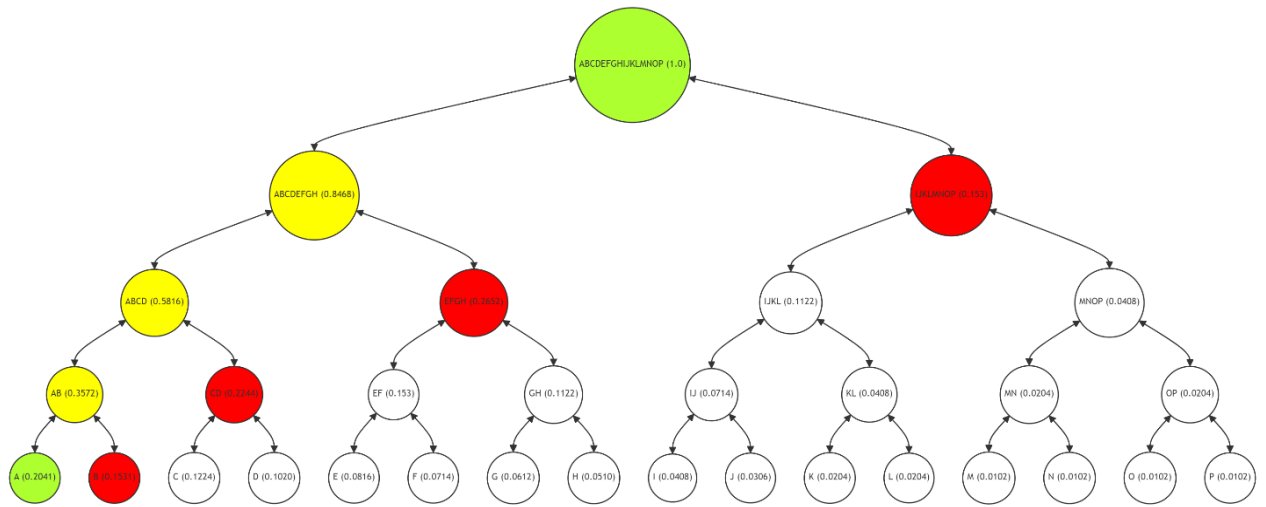

**Figure 4:** Merkle Path (red nodes B, CD, EFGH, IJKLMNOP) for Leaf Node A

In Figure 5, the Merkle Path for verifying leaf node G (with a frequency of 0.0612) is shown. The path (nodes H, EF, ABCD, IJKLMNOP) is marked in red, with node G and the root in green, and the intermediate nodes (GH, EFGH, ABCDEFGH) in yellow. The verification process for G follows the same principle as for A, recalculating hashes along the red path to validate the data's integrity.

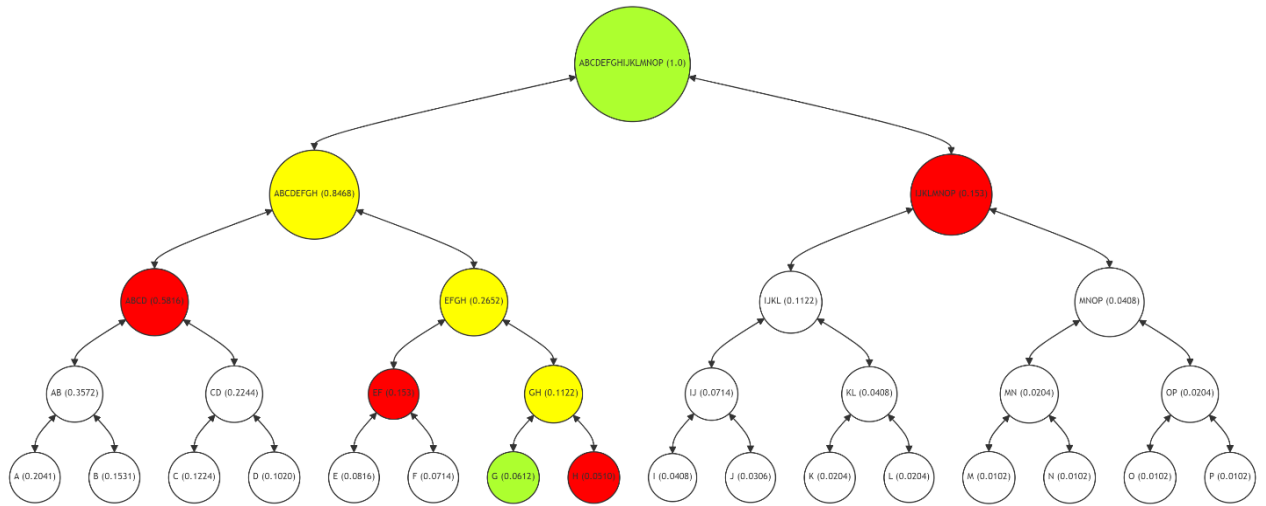

**Figure 5:** Merkle Path (red nodes H, EF, ABCD, IJKLMNOP) for Leaf Node G

The Figure 6 demonstrates the Merkle Path for leaf node P (with a frequency of 0.0102), with the path (nodes O, MN, IJKL, ABCDEFGH) in red, P and the root in green, and intermediate nodes (OP, MNOP, IJKLMNOP) in yellow. The process for verifying P's integrity mirrors that of A and G, emphasizing the consistent approach across the tree.

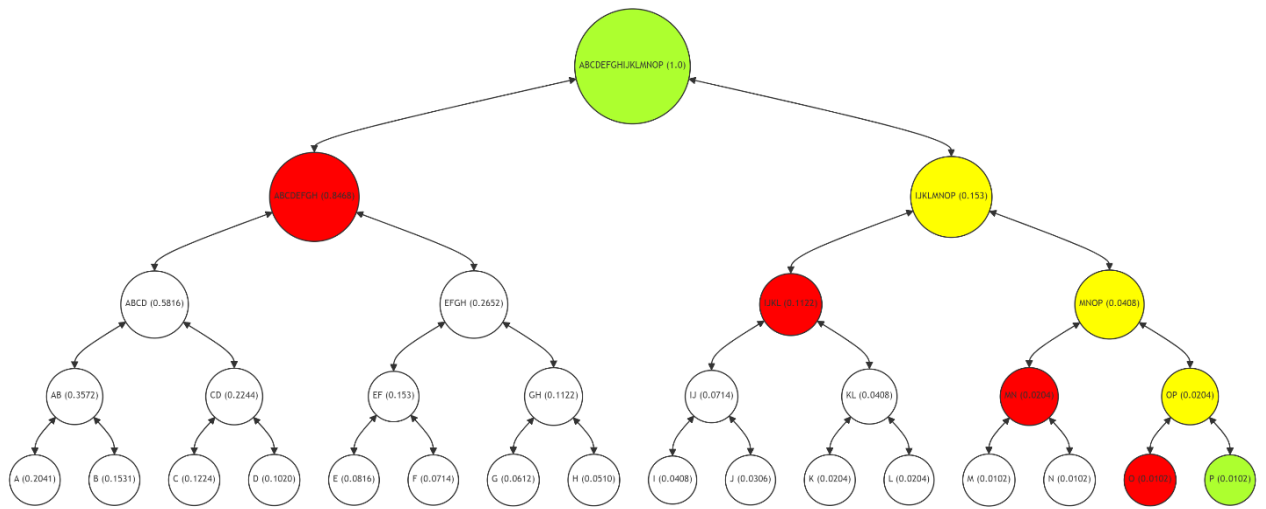

**Figure 6:** Merkle Path (red nodes O, MN, IJKL, ABCDEFGH) for Leaf Node P

This consistency in verification, while ensuring uniform security and integrity checks, does not account for the varying frequencies of data access and modification. It leads to a rigid and sometimes inefficient system, especially in a dynamic environment like blockchain, where data access patterns can vary significantly.

Thus, the current Merkle Tree verification process, as illustrated in these figures, is a rather primitive and blunt approach. It treats all data equally, irrespective of its usage frequency, leading to potential inefficiencies in computational resources. Our proposed solution aims to revolutionize this process by introducing adaptive Merkle Trees. These trees will optimize verification paths based on data usage frequency, significantly reducing the complexity and cost of verifying frequently accessed data. This innovative approach promises to enhance the efficiency and scalability of blockchain systems, tailoring the verification process to the dynamic needs of the network. By differentiating between frequently and infrequently accessed data, adaptive Merkle Trees can allocate computational resources more effectively, ensuring faster and more cost-efficient data verification. This method not only optimizes the blockchain's performance but also aligns with the evolving nature of blockchain usage, where certain data nodes may become hotspots of activity.

## 3. Our Idea for Optimizing Trees in Blockchain

Figure 7 represents an innovative adaptation of the traditional Merkle Tree, incorporating principles of Shannon-Fano and Huffman statistical coding. Unlike the balanced Merkle Tree, this adaptive structure is intentionally unbalanced to optimize the verification process based on the frequency of data usage. Each leaf node (A-P) still represents a block of data with a unique hash value, but their placement in the tree now correlates with the probability of their usage.

In this adaptive Merkle Tree, the most frequently used data nodes (A, B, C, D) are positioned closer to the root, significantly shortening the path required for their verification. This strategic placement reduces the computational complexity and time required for verifying frequently accessed data. Conversely, less frequently used data nodes (M, N, O, P) are placed further from the root, reflecting their lower probability of access.

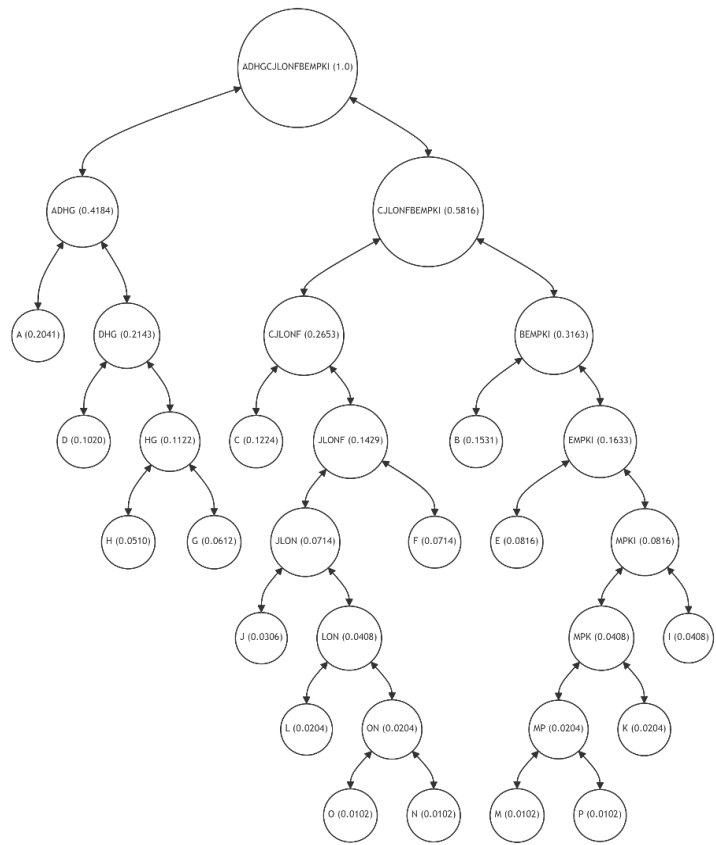

**Figure 7:** Adaptive Merkle Tree

The structure of this tree is a direct application of Shannon-Fano and Huffman coding principles, where the most common elements are given shorter codes (or paths in the case of a Merkle Tree). This approach ensures that the average path length for verification is minimized, aligning the computational effort with the actual usage patterns of the data within the blockchain.

In the Figure 8, the Merkle Path for leaf node A (highlighted in green) is significantly shorter than in a balanced Merkle Tree. The path (marked in red) includes nodes DHG and CJLONFBEMPKI, with intermediate calculations (in yellow) at node ADHG. This optimized path reflects the high frequency of usage for node A, making the verification process faster and more cost-effective. The integrity of node A can be verified with fewer computational steps, demonstrating the efficiency of the adaptive Merkle Tree in handling frequently used data.

For leaf node G (Figure 9), the Merkle Path includes nodes H, D, and CJLONFBEMPKI, with intermediate calculations at nodes HG and ADHG. This path, while longer than that for node A, is still optimized based on the usage frequency of G. The adaptive tree structure ensures that the verification process remains efficient, even for nodes with moderate usage. This approach balances the need for data integrity with computational efficiency, tailoring the verification complexity to the usage pattern of each node.

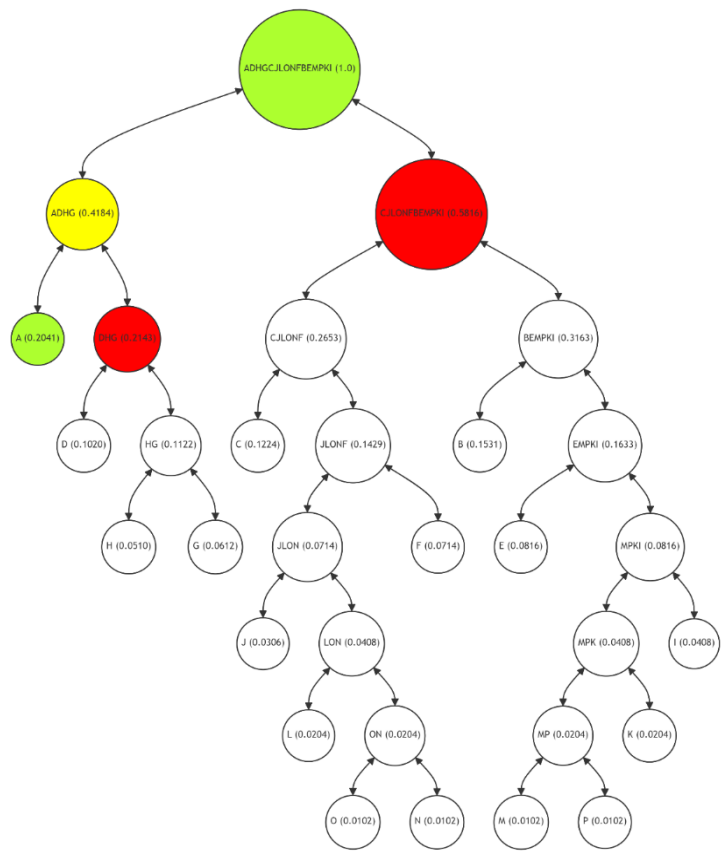

**Figure 8:** Optimized Merkle Path for High-Frequency Leaf Node A

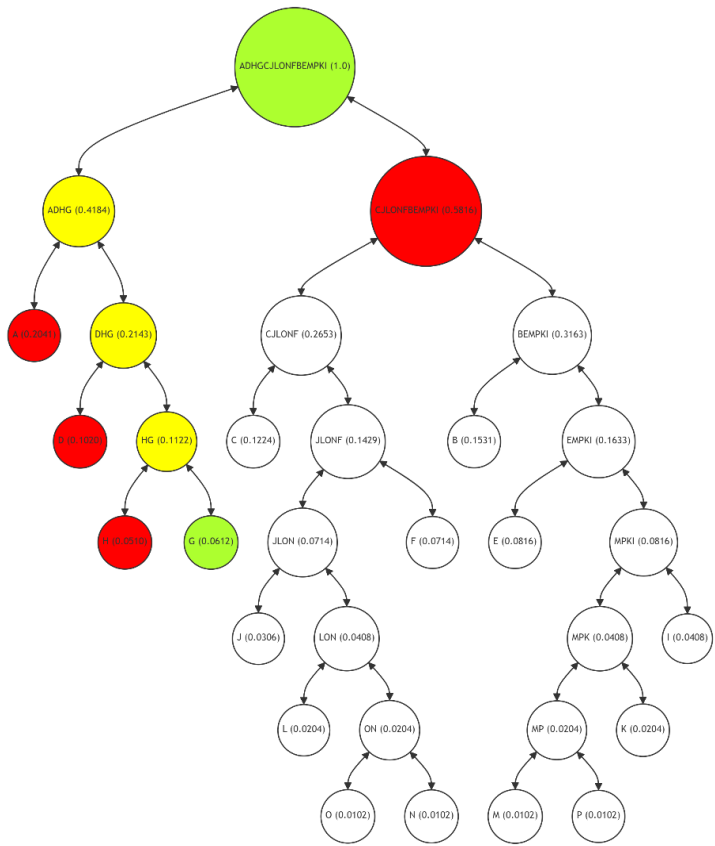

**Figure 9:** Adaptive Merkle Path for Moderately Used Leaf Node G

The Merkle Path for leaf node P (Figure 10), a less frequently used node, is longer, including nodes M, K, I, E, B, CJLONF, and ADHG. The path reflects P's lower usage frequency, with more intermediate calculations (nodes MP, MPK, MPKI, EMPKI, BEMPKI, and CJLONFBEMPKI) required for verification. While this makes the verification process for P more resource-intensive, it is justified by the node's infrequent use. This example illustrates how the adaptive Merkle Tree allocates computational resources more efficiently, focusing on optimizing the paths for more frequently used nodes.

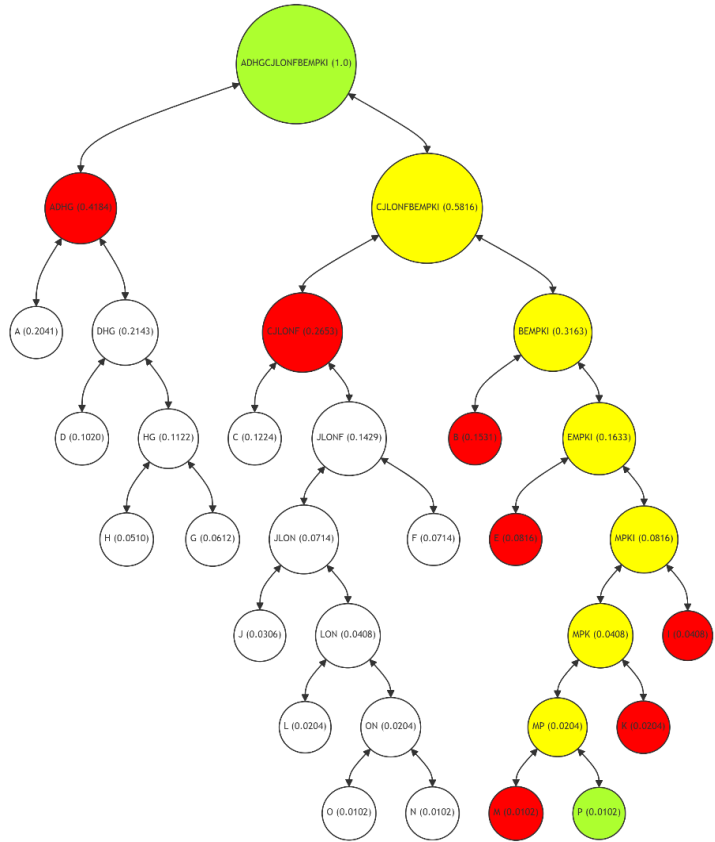

**Figure 10:** Extended Merkle Path for Low-Frequency Leaf Node P

Thus, the adaptive Merkle Tree approach significantly enhances the efficiency of data verification in blockchain systems. For high-frequency nodes like A, the verification process is streamlined, requiring fewer computational steps and resources. This optimization can lead to a verification process that is up to twice as fast and cost-effective compared to a balanced Merkle Tree. Conversely, for nodes with lower usage frequencies, like P, the longer verification path is a reasonable trade-off, considering their infrequent access.

**4. Efficiency of adaptive Merkle trees**

In this work, we delve into the comparative complexity of data integrity verification between the conventional balanced Merkle Tree and the proposed adaptive Merkle Tree model. The balanced Merkle Tree's average path length is determined by

$$k \approx \log_m n,$$

where $m$ represents the maximum allowable number of child nodes per node (the arity of the tree), and $n$ is the count of unique symbols within the alphabet.

Conversely, the adaptive Merkle Tree's average path length mirrors the average length of a Huffman code, calculated as the weighted sum of all code lengths, with the probabilities of the corresponding symbols serving as weights:

$$k_A = \sum_{i=1}^{n} p_i \times l_i,$$

where $p_i$ is the probability of the $i$ th symbol, and $l_i$ is the length of the code for the $i$ th symbol.

The theoretical minimum average length of a Huffman code, given a specific probability distribution, can be derived from the entropy formula:

$$k_A \geq H = -\sum_{i=1}^{n} p_i \times \log_m(p_i).$$

Thus, the efficiency of a Huffman code increases as its average code length approaches the entropy of the distribution.

For the binary tree example ($m = 2$) discussed, the Huffman code's average length is approximately 3.49 bits per symbol, closely approximating the entropy of the symbol probabilities distribution, which is about 3.46 bits per symbol. These figures suggest that the Huffman code from our example is remarkably close to the theoretical minimum average code length defined by entropy. Ideally, if the code were perfectly optimal, its average length would equal the entropy.

Transforming these assessments into a comparison of the complexity of data integrity verification in both the classical balanced and the proposed adaptive Merkle Tree yields:

- For a balanced binary tree, the average Merkle path length is $k \approx 4$;

- For an adaptive binary Merkle Tree, the average path length is $k_A \approx 3.49$, indicating an efficiency gain of approximately 13%. This gain is reflected in the reduced average number of hash computations required for verifying the integrity of leaf data.

The efficiency gain increases with the growing disparity between the probabilities of leaf data. In the extreme case, where one leaf has a 100% probability and all others have 0%, the maximum efficiency gain—up to 100%—can be observed. Although this represents a hypothetical scenario, it is intriguing to model real adaptive Merkle Trees, including non-binary types, and assess the effectiveness of our proposed solution. In Ethereum, Patricia trees are utilized, and our aim is to extend our approach to this case as well. Furthermore, algorithms for the gradual restructuring of balanced trees into an unbalanced form are of particular interest. We propose a protocol for such gradual restructuring, which utilizes newly added nodes to replace high-frequency nodes in the existing tree. These high-frequency nodes are relocated within the tree to positions that correspond to their usage probability, allowing us to incrementally modify the tree's configuration and enhance the efficiency of blockchain integrity checks without a complete overhaul.

## 5. Algorithm for Merkle Tree Restructuring

The restructuring of a Merkle Tree, aimed at optimizing its efficiency for blockchain applications, necessitates adherence to two primary criteria:

- Minimization of Average Path Length: The restructuring process must account for the usage frequency of each leaf, ensuring that the average path length, $k_A$, approaches the theoretical minimum, or the average entropy, $H$. The deviation between $k_A$ and $H$ is assessed through the average discrepancy:

$$\Delta = k_A - H \quad (1)$$

with each elemental discrepancy defined as:

$$\Delta_i = p_i(l_i + \log_m(p_i)), \quad (2)$$

where

$$\Delta = \sum_{i=1}^{n} \Delta_i .$$

This requirement mandates the availability of a list of probabilities, $p_i$, and path lengths, $l_i$, for each leaf during the restructuring process, updating only as necessary.

- Minimization of Altered Paths: The algorithm should limit modifications to a minimal subset of nodes, reflecting the reality that only a few accounts are activated in any given transaction, including complex smart transactions. This approach ensures that inactive accounts retain their positions and paths within the tree, preserving the integrity of user data and access pathways. To adhere to this criterion, the algorithm must maintain a list of leaves (nodes) eligible for restructuring, focusing solely on those affected by current transactions.

**Restructuring Algorithm (A Single Iteration)**

Input:

- A tree (or tree fragment) with its root, intermediate nodes, and leaves (the bottom layer of the tree nodes).
- The probability distribution (frequencies) of the tree's leaves.
- A set (list) of leaves available for restructuring.
- A new leaf and/or a new probability distribution for all tree leaves.

Output:

- A restructured tree (or tree fragment) optimized according to the criterion of minimizing the average discrepancy ($\Delta$).

Algorithm Steps:

- Utilize the set (list) of leaves available for restructuring to formulate all possible restructuring alternatives for the tree (or tree fragment).
- Evaluate the average discrepancy ($\Delta$) for each alternative.
- Select the alternative with the lowest average discrepancy ($\Delta$).
- Adopt the selected alternative as the algorithm's output.

The most challenging aspect of this algorithm is Step 1, which involves generating all possible restructuring alternatives for the tree. This process is crucial for identifying the most efficient tree configuration that minimizes the average path length while accommodating the dynamic nature of blockchain transactions. To demonstrate the algorithm's functionality amidst the increasing number of alternatives, several illustrative examples will be provided, showcasing its application in various scenarios.

## 6. Examples of Merkle Tree Restructuring Algorithm Execution

To illustrate the algorithm's application, let's consider the case of a binary tree where one new leaf is added at each iteration. Initially, we form a list of alternatives for all leaves in the previous tree configuration. Each leaf can be transformed into an intermediate node with two child nodes: one from the previous configuration and one new (added) leaf.

### 6.1 Example 1: Restructuring a Binary Tree by Adding One Leaf

Suppose we have a small binary ($m=2$) balanced tree consisting of two nodes A and B, with probabilities 7/8 and 1/8, respectively (see Fig. 11, a).

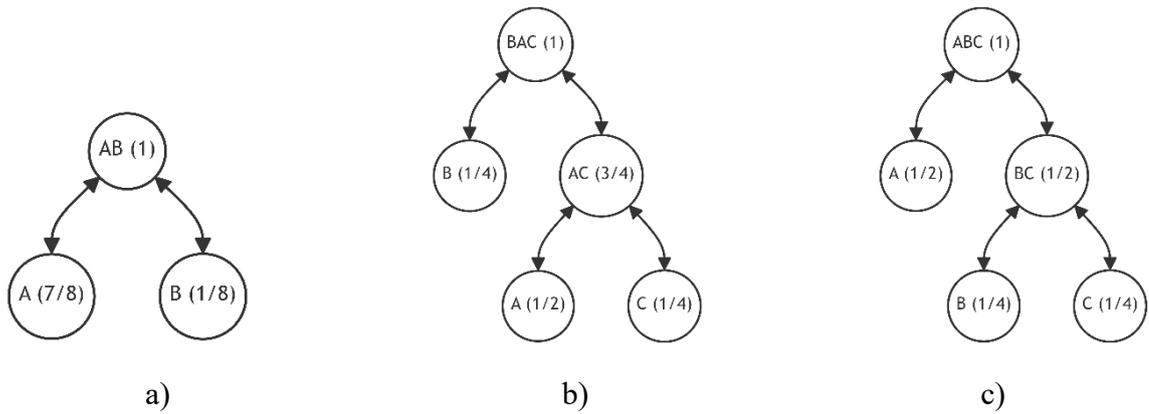

Figure 11: Binary Tree Restructuring (First Iteration)

Assume that at the next moment, a new leaf C is created with probabilities now equal to:

A (1/2), B (1/4), C (1/4).

Our goal is to add this new leaf C in such a way as to minimize the average discrepancy (1). Here and subsequently, we assume that all branches from the previous tree configuration are available for addition.

### 6.1.1. First Iteration

On the first iteration, we have two alternatives for adding the new leaf to the previous tree configuration. These alternatives are presented in Fig. 11 and Table 1.

The first alternative (see Fig. 11, b) corresponds to adding node C (1/4) to the branch with node A (1/2). As we can see from Table 1, this increases the discrepancy $\Delta$. The second alternative (see Fig. 11, c) is more preferable as the discrepancy (1) here is significantly lower (equals zero), i.e., node C (1/4) should be added to the branch with node B (1/4).

Table 1. Discrepancy Values for Two Alternative Ways of Tree Restructuring

| | First alternative, Fig. 11, b | | | | | | Second alternative, Fig. 11, c | | | | |
|---|---|---|---|---|---|---|---|---|---|---|---|
| | $p_i$ | $l_i$ | $p_i l_i$ | $-p_i \log_2(p_i)$ | $\Delta_i$ | | | $p_i$ | $l_i$ | $p_i l_i$ | $-p_i \log_2(p_i)$ | $\Delta_i$ |
| A | ½ | 2 | 1 | ½ | ½ | | A | ½ | 1 | ½ | ½ | 0 |
| B | ¼ | 1 | ¼ | ½ | -¼ | | B | ¼ | 2 | ½ | ½ | 0 |
| C | ¼ | 2 | ½ | ½ | 0 | | C | ¼ | 2 | ½ | ½ | 0 |
| | | | $\Delta$ | | ½ | | | | | $\Delta$ | | 0 |

Therefore, by the criterion of minimizing (1), we select the second alternative, i.e., the tree presented in Fig. 11, c. From the perspective of path length, this option is optimal as its average discrepancy (1) equals zero. Essentially, this indicates that we have achieved an ideal structure for this probability distribution.

Continuing from the initial iteration of the Merkle Tree restructuring algorithm, let us delve into subsequent iterations to further elucidate the process and its outcomes.

### 6.1.2. Second Iteration

Let us assume that in the second iteration, a new leaf, D, is introduced, leading to a new probability distribution among the leaves:

A (1/2), B (1/8), C (1/4), and D (1/8).

Building upon the tree's previous state (refer to Fig. 11, c), we are presented with three distinct restructuring alternatives (illustrated in Fig. 12):

a) Integrating the new node D (1/8) into the branch containing node A (1/2);
b) Integrating the new node D (1/8) into the branch containing node B (1/8);
c) Integrating the new node D (1/8) into the branch containing node C (1/4).

For each alternative, we calculate the average discrepancy, as detailed in Table 2. The most favorable alternative, marked by a zero discrepancy, is highlighted in the table.

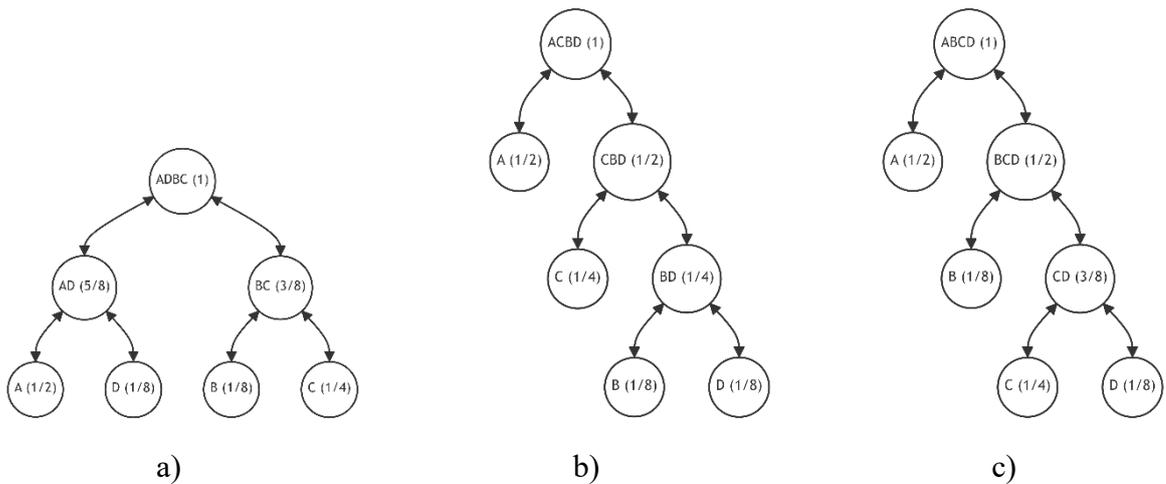

a) b) c)

**Figure 12:** Binary Tree Restructuring (Second Iteration)

Table 2. Comparison of Alternatives (Second Iteration)

|  | $k_A$ | $H$ | $\Delta$ |
|---|---|---|---|
| Fig. 12, a | 2 | 1.75 | 0.25 |
| Fig. 12, b | 1.75 | 1.75 | 0 |
| Fig. 12, c | 1.875 | 1.75 | 0.125 |

Accordingly, the second alternative (see Fig. 12, b) emerges as the most preferable, characterized by a zero average discrepancy, indicating an optimal restructuring choice under the given criteria.

### 6.1.3. Third Iteration

Advancing to the third iteration, let's hypothesize the addition of another new leaf, E, resulting in the following probability distribution:

A (1/2), B (1/8), C (1/8), D (1/8), E (1/8).

The tree structure that achieves a zero discrepancy, indicative of an optimal configuration minimizing the average path length, is depicted in Fig. 13.

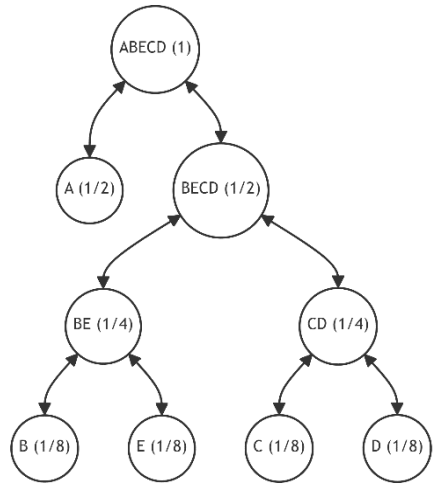

**Figure 13:** Binary Tree Restructuring (Third Iteration)

Achieving a zero discrepancy signifies that we have attained an optimal tree structure, effectively minimizing the average path length.

### 6.1.4. Fourth Iteration

During the fourth iteration, we face the task of incorporating an additional leaf, resulting in a new probability distribution:

A (1/2), B (1/4), C (1/16), D (1/16), E (1/16), F (1/16).

Ideally, a tree structure with a zero discrepancy ($\Delta = 0$) would align with the configuration depicted in Figure 14.a. However, this ideal structure cannot be achieved by simply adding a leaf to the previous configuration (as shown in Figure 13).

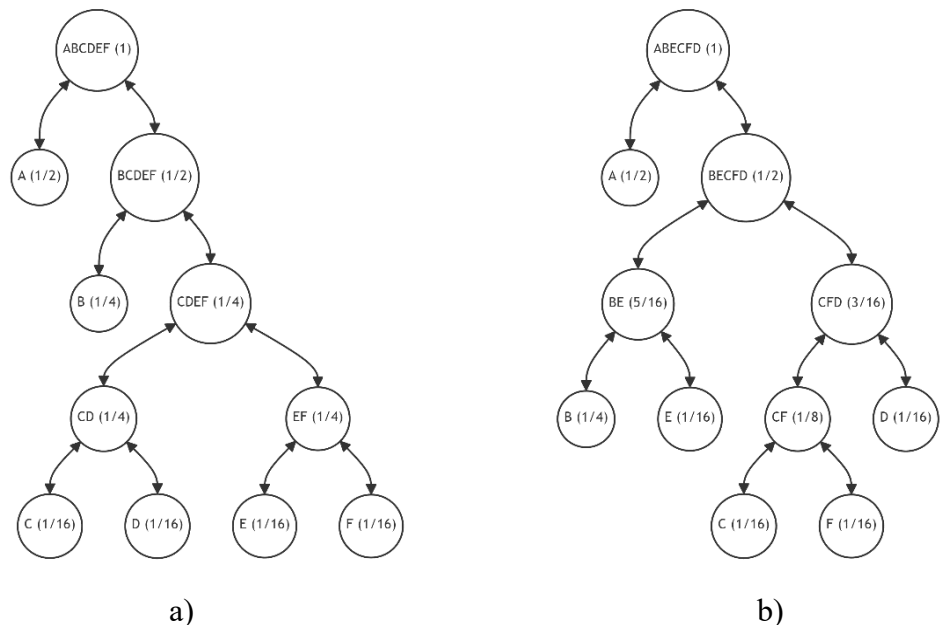

a)          b)

**Figure 14:** Binary Tree Restructuring (Fourth Iteration)

Utilizing the tree from Figure 13, we identify five potential alternatives (refer to Table 3), none of which lead to the desired configuration seen in Figure 14.a.

In Table 3, we present a comparison of all possible alternatives based on the average discrepancy value (1). It becomes evident that the last three alternatives are equivalent in terms of their potential outcomes, and thus, the choice of restructuring can be made arbitrarily. We decide to employ a lexicographical ordering rule, selecting alternative "C" as illustrated in Figure 14.b. Although this new structure is not optimal, it achieves the minimum average discrepancy of 0.125 among all possible restructuring scenarios.

Table 3. Comparison of Alternatives (Fourth Iteration)

|  | $k_A$ | H | Δ |
|---|---|---|---|
| Restructuring Branch with Leaf A | 2.4375 | 2 | 0.4375 |
| Restructuring Branch with Leaf B | 2.3125 | 2 | 0.3125 |
| Restructuring Branch with Leaf C | 2.125 | 2 | 0.125 |
| Restructuring Branch with Leaf D | 2.125 | 2 | 0.125 |
| Restructuring Branch with Leaf E | 2.125 | 2 | 0.125 |

### 6.1.5. Iterations 5-10

For further illustration, let's assume that each subsequent iteration introduces one additional leaf, with the probability distribution for the tree's leaves as specified in Table 4. This table also indicates the number of available restructuring alternatives and (in parentheses) the number of alternatives with the minimum discrepancy value. The last column provides the minimum discrepancy value (11) across all alternatives. Figures 15-20 showcase the restructuring outcomes based on the chosen optimal alternative.

Table 4. List of Leaves and Their Probabilities per Iteration, Number of Restructuring Alternatives, and Minimum Discrepancy Value

| Iteration number | List of Leaves and Their Probabilities | Number of Restructuring Alternatives | Minimum Discrepancy Value |
|---|---|---|---|
| 5 | A (1/2), B (1/4), C (1/16), D (1/16), E (1/32), F (1/16), G (1/32) | 6 (1) | 0.125 |
| 6 | A (1/4), B (1/4), C (1/16), D (1/8), E (1/16), F (1/8), G (1/16), H (1/16) | 7 (6) | 0.25 |
| 7 | A (1/4), B (1/4), C (1/16), D (1/8), E (1/32), F (1/8), G (1/32), H (1/16), I (1/16) | 8 (1) | 0.1875 |
| 8 | A (1/4), B (1/4), C (1/16), D (1/8), E (1/32), F (1/16), G (1/32), H (1/16), I (1/16), J (1/16) | 9 (2) | 0.125 |
| 9 | A (1/8), B (1/4), C (1/16), D (1/8), E (1/16), F (1/8), G (1/32), H (1/16), I (1/16), J (1/16), K (1/32) | 10 (2) | 0.185 |
| 10 | A (1/8), B (1/4), C (1/16), D (1/8), E (1/32), F (1/8), G (1/32), H (1/16), I (1/16), J (1/16), K (1/32), L (1/32) | 11 (2) | 0.185 |

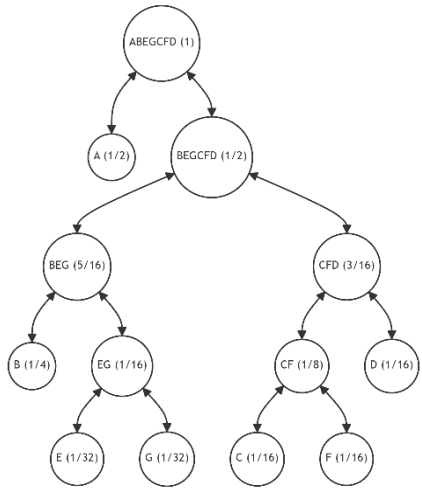

**Figure 15:** Binary Tree Restructuring (Iterations 5)

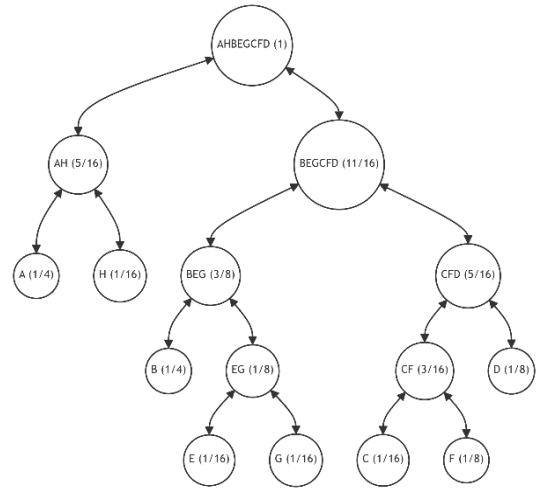

**Figure 16:** Binary Tree Restructuring (Iterations 6)

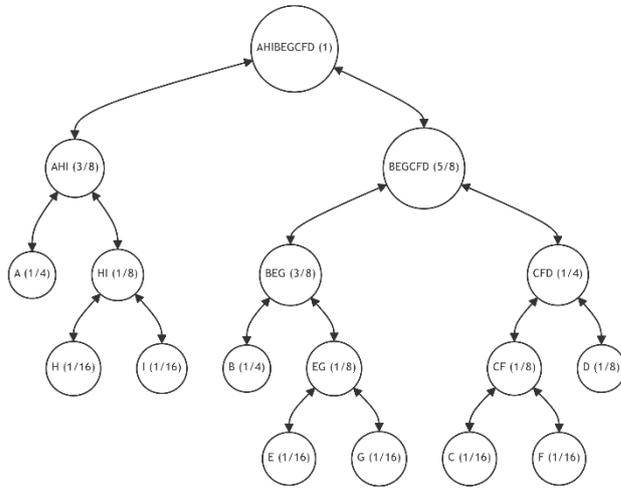

**Figure 17:** Binary Tree Restructuring (Iterations 7)

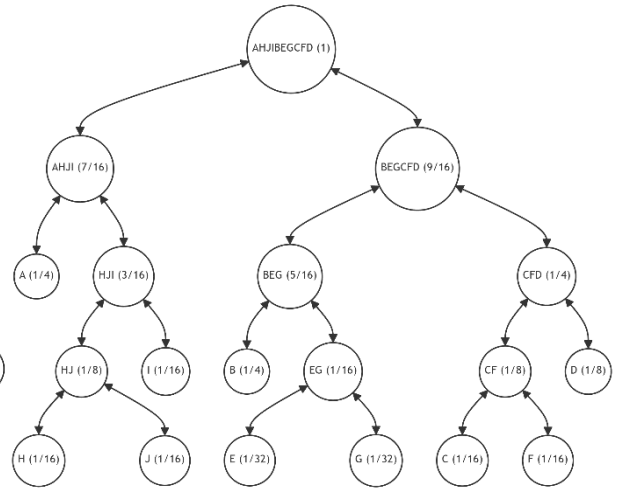

**Figure 18:** Binary Tree Restructuring (Iterations 8)

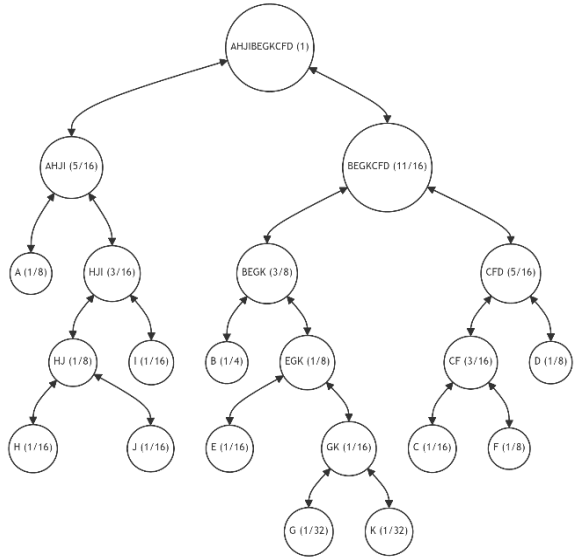

**Figure 19:** Binary Tree Restructuring (Iterations 9)

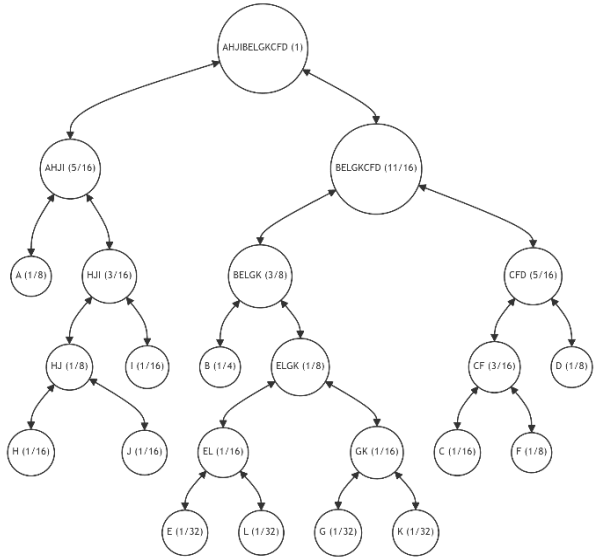

**Figure 20:** Binary Tree Restructuring (Iterations 10)

The table reveals that, despite an increase in the number of alternatives with each iteration, the number of most preferable restructuring options remains limited. Moreover, despite fluctuations in the probabilities of individual leaves, we consistently approach an optimal tree structure.

To further enhance the efficiency of tree restructuring, we propose expanding the algorithm's parameters. This involves forming potential alternatives not just by adding leaves but by considering various scenarios for swapping the positions of leaf pairs. This approach is particularly relevant in blockchain transactions, which typically involve at least two accounts, thus allowing for the repositioning of leaves by exchanging their locations.

This expansion of the algorithm's capabilities demonstrates our commitment to optimizing tree structures for improved verification efficiency, paving the way for more dynamic and efficient blockchain architectures.

**6.2. Example 1.1: Binary Tree Restructuring Through Leaf Node Swapping**

To demonstrate the algorithm's functionality, let's revisit the outcome of the sixth iteration from the previous example, which represented the worst case in terms of average discrepancies, i.e., it was the most suboptimal structure depicted in Figure 16. We now aim to improve this structure by swapping the positions of leaf pairs to minimize the discrepancy ($\Delta$).

Instead of considering all possible leaf pairs, we focus only on those leaves whose elementary discrepancies ($\Delta_i$) are non-zero. Essentially, the criterion $\Delta_i \neq 0$ indicates that the $i$-th leaf is "out of place." For the graph in Figure 16, we have:

- Leaves: A, B, C, D, E, F, G, H;
- Probabilities ($p_i$): 0.25, 0.25, 0.06, 0.13, 0.06, 0.13, 0.06, 0.06;
- Path Lengths ($l_i$): 2, 3, 4, 3, 4, 4, 4, 2;
- Discrepancies ($\Delta_i$): 0, 0.25, 0, 0, 0, 0.125, 0, -0.125.

Thus, leaves B, F, and H are candidates for swapping positions, yielding three alternatives:

- Swapping B and F results in $\Delta = 0.375$;
- Swapping B and H results in $\Delta = 0.0625$;
- Swapping F and H results in $\Delta = 0.125$.

Clearly, the best alternative for our example is to swap leaves B and H. Visually, this corresponds to the same graph shown in Figure 16.1 – identical to Figure 16 but with B and H swapped.

After this restructuring, we observe the following distribution of elementary discrepancies:

- Leaves: A, H, C, D, E, F, G, B;
- Probabilities ($p_i$): 0.25, 0.0625, 0.0625, 0.125, 0.0625, 0.125, 0.0625, 0.25;
- Path Lengths ($l_i$): 2, 3, 4, 3, 4, 4, 4, 2;
- Discrepancies ($\Delta_i$): 0, -0.0625, 0, 0, 0, 0.125, 0, 0.

Now, the only viable alternative is to swap leaves H and F, which results in a zero average discrepancy, indicating an optimal structure in terms of minimizing the average path length in the binary tree. The outcome of this optimization is depicted in Figure 16.2.

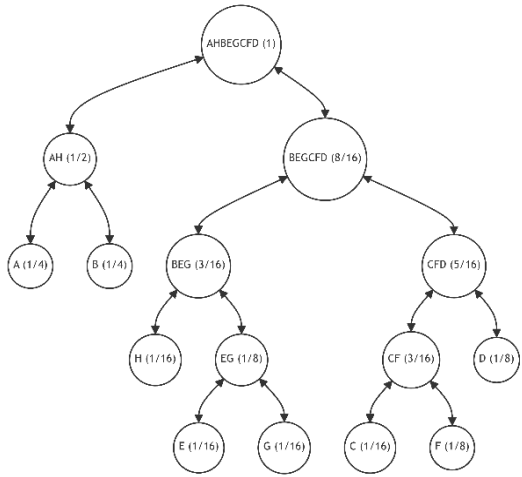
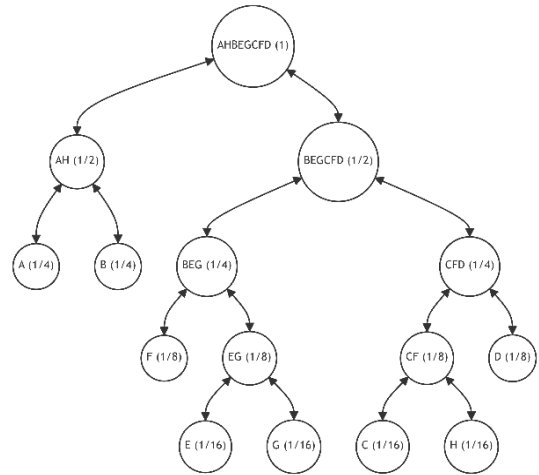

**Figure 16.1:** Graph Optimization Post-Sixth Iteration (Swapping Leaves B and H)

**Figure 16.2:** Graph Optimization Post-Sixth Iteration (Swapping Leaves H and F)

It's important to note that non-binary trees are often used in practical scenarios. For instance, Ethereum's blockchain utilizes a Patricia-Merkle tree, which can have up to 16 child nodes, i.e., $m=16$.

Let's demonstrate the algorithm's operation on a simple example of a non-binary ($m=4$) tree.

### 6.3. Example 2.1: Restructuring a Non-Binary Tree by Adding a Single Leaf

In this example, we explore a non-binary tree where each node can have up to four children ($m=4$), closely mirroring a simplified real-world scenario of Patricia-Merkle trees in the Ethereum blockchain.

Let's assume the initial state of the tree is as shown in Figure 11.a, similar to the previous example. Also, let the newly added leaves and their probability distributions follow the pattern established in Example 1. These changes in probabilities are summarized in Table 5, which also lists the number of alternatives and the unique discrepancy value (1). Figures 21-30 depict the corresponding tree graphs.

Table 5. List of Leaves and Their Probabilities per Iteration, Number of Restructuring Alternatives, and Minimum Discrepancy Value

| Iteration Number | List of Leaves and Their Probabilities | Number of Alternatives | Minimum Discrepancy Value (1) |
|---|---|---|---|
| 1 | A (1/2), B (1/4), C (1/4) | 3 (1) | 0. 25 |
| 2 | A (1/2), B (1/8), C (1/4), D (1/8) | 4 (1) | 0.125 |
| 3 | A (1/2), B (1/8), C (1/8), D (1/8), E (1/8) | 4 (3) | 0.25 |
| 4 | A (1/2), B (1/4), C (1/16), D (1/16), E (1/16), F (1/16) | 6 (1) | 0.375 |

| 5 | A (1/2), B (1/4), C (1/16), D (1/16), E (1/32), F (1/16), G (1/32) | 7 (1) | 0.34375 |
| 6 | A (1/4), B (1/4), C (1/16), D (1/8), E (1/16), F (1/8), G (1/16), H (1/16) | 7 (1) | 0.25 |
| 7 | A (1/4), B (1/4), C (1/16), D (1/8), E (1/32), F (1/8), G (1/32), H (1/16), I (1/16) | 9 (1) | 0.21875 |
| 8 | A (1/4), B (1/4), C (1/16), D (1/8), E (1/32), F (1/16), G (1/32), H (1/16), I (1/16), J (1/16) | 10 (1) | 0.15625 |
| 9 | A (1/8), B (1/4), C (1/16), D (1/8), E (1/16), F (1/8), G (1/32), H (1/16), I (1/16), J (1/16), K (1/32) | 10 (1) | 0.21875 |
| 10 | A (1/8), B (1/4), C (1/16), D (1/8), E (1/32), F (1/8), G (1/32), H (1/16), I (1/16), J (1/16), K (1/32), L (1/32) | 12 (1) | 0.21875 |

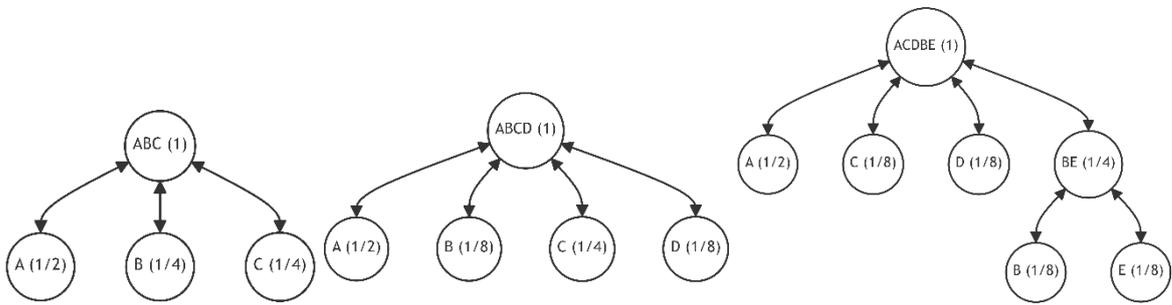

**Figure 21:** Tree Restructuring (Iterations 1)   **Figure 22:** Tree Restructuring (Iterations 2)   **Figure 23:** Tree Restructuring (Iterations 3)

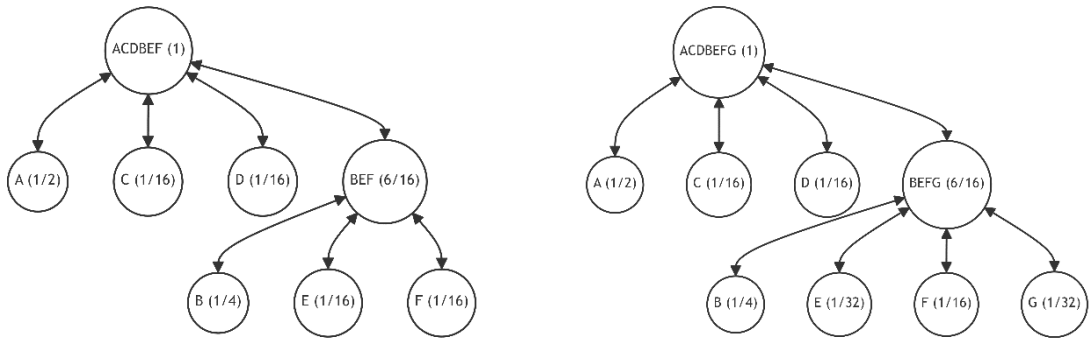

**Figure 24:** Tree Restructuring (Iterations 4)   **Figure 25:** Tree Restructuring (Iterations 5)

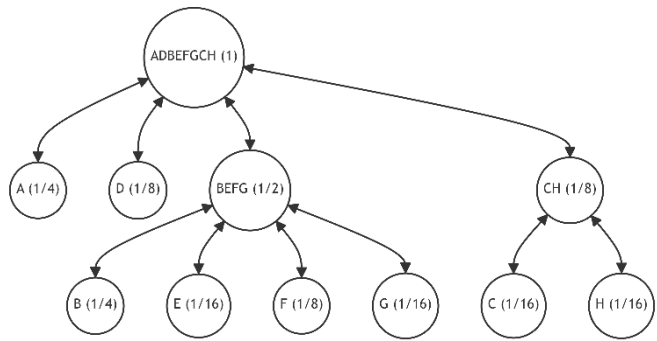

**Figure 26:** Tree Restructuring (Iterations 6)

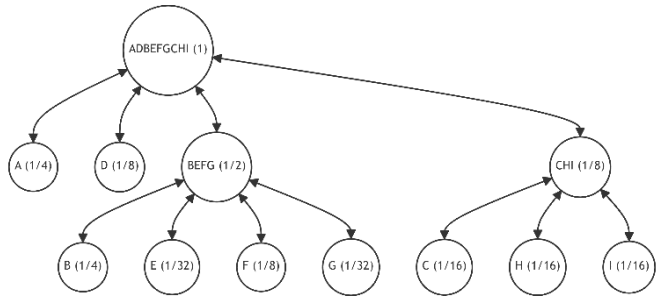

**Figure 27:** Tree Restructuring (Iterations 7)

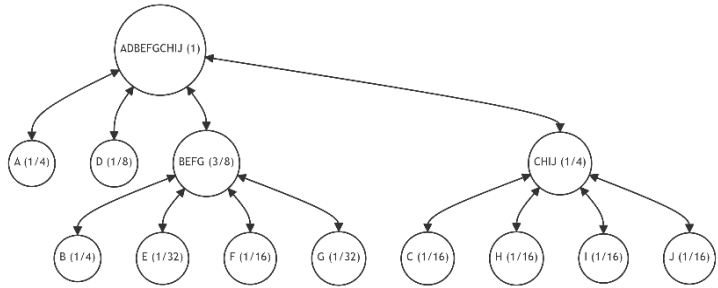

**Figure 28:** Tree Restructuring (Iterations 8)

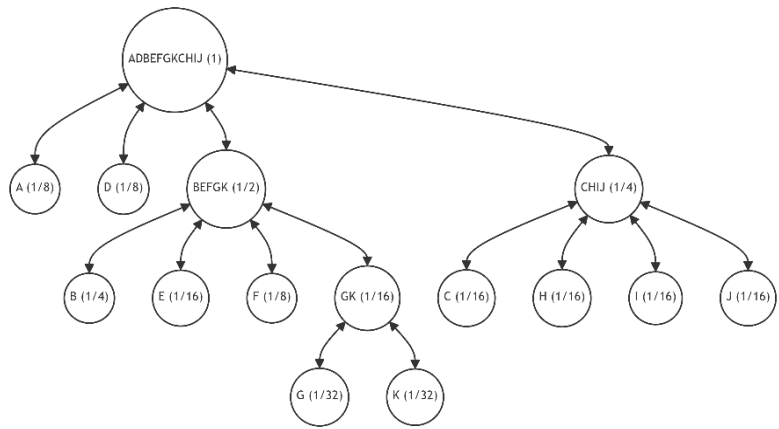

**Figure 29:** Tree Restructuring (Iterations 9)

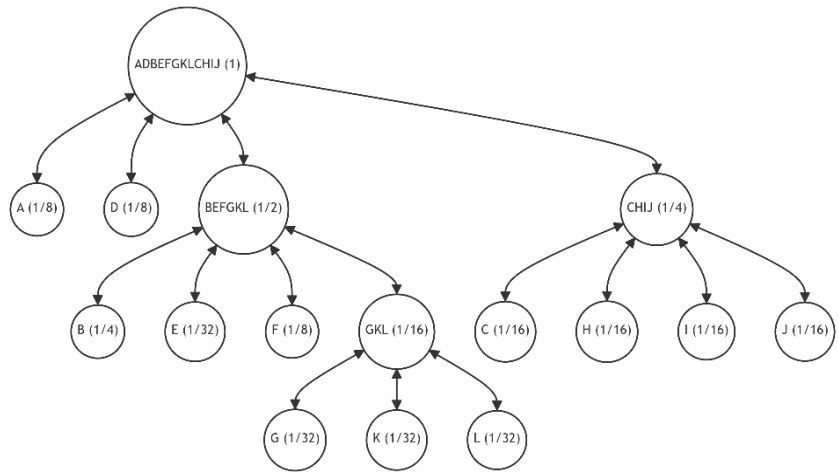

**Figure 30:** Tree Restructuring (Iterations 10)

The calculations of discrepancies in Table 5 show that the configuration of the restructured trees tends towards optimality by minimizing the average path length.

Now, let's demonstrate the algorithm's operation in the mode of swapping positions between pairs of nodes, as in Example 1.1.

### 6.4. Example 2.2: Restructuring a Non-Binary Tree Through Leaf Pair Swapping

Let's delve into the most challenging scenario from Example 2, which resulted in the highest discrepancy value. This scenario corresponds to the tree graph obtained after the fourth iteration, as depicted in Figure 24. We will demonstrate how swapping the positions of leaf pairs can enhance this structure.

To form a set of alternatives for the graph in Figure 24, we observe:

- Leaves: A, B, C, D, E, F;
- Probabilities ($p_i$): 0.50, 0.25, 0.06, 0.06, 0.06, 0.06;
- Path Lengths ($l_i$): 1, 2, 1, 1, 2, 2;
- Discrepancies ($\Delta_i$): 0.25, 0.25, -0.0625, -0.0625, 0, 0.

Focusing on pairs with differing path lengths, we identify:

- Swapping positions between leaves A and B, Δ=0.625;
- Swapping positions between leaves B and C, Δ=0.1875;
- Swapping positions between leaves B and D, Δ=0.1875.

The last two alternatives are equivalent and halve the discrepancy (1), thus optimizing the final tree structure (see Figure 24.1).

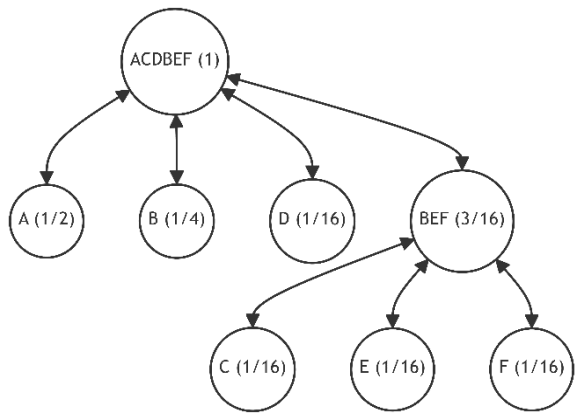

**Figure 24.1:** Graph Optimization After the Fourth Iteration (Swapping Positions Between Leaves B and C)

This approach allows for the dynamic restructuring of trees, minimizing the divergence between the current and optimal graph structures. By combining different rules (adding new leaves and swapping positions of existing leaves), we can achieve highly efficient structures that minimize the average path length.

Through this methodology, we underscore the algorithm's capability to swiftly adapt tree structures, ensuring an optimal configuration that aligns closely with the theoretical minimum discrepancy. This adaptability is crucial for maintaining efficient data verification processes in blockchain technologies, where the dynamic nature of transactions necessitates a flexible yet robust system for ensuring data integrity.

The proposed algorithm exemplifies a significant advancement in optimizing tree structures for blockchain applications, particularly in scenarios where non-binary trees, such as Patricia-Merkle trees used in Ethereum, are prevalent. By judiciously applying leaf swapping and addition strategies, we can significantly enhance the efficiency of these cryptographic structures, paving the way for more scalable and cost-effective blockchain operations.

In conclusion, let's explore another example of a tree with $m=16$, which can be considered as restructuring a fragment of the Patricia-Merkle tree in the Ethereum blockchain.

## 6.5. Example 2.3: Restructuring a Patricia-Merkle Tree Fragment Through Leaf Pair Swapping

Imagine we have a fragment of the Patricia-Merkle tree with leaves (and probabilities) assigned as follows (see Figure 31):

A (0.003906), B (0.0625), C (0.16529), D (0.0625), E (0.0625), F (0.0625), G (0.0625), H (0.0625), I (0.003906), J (0.0625), K (0.0625), L (0.0625), M (0.0625), N (0.003906), O (0.0625), P (0.000244), Q (0.0625), R (0.01), S (0.0625), T (0.000244).

For this tree configuration, we have an average discrepancy Δ=0.1297. By swapping the positions of leaves A (0.003906) and E (0.0625), we achieve the lowest discrepancy Δ=0.0711 among all possible alternatives. This tree is depicted in Figure 32.

Continuing to apply the algorithm, we swap the positions of leaves P (0.000244) and R (0.01), resulting in a discrepancy Δ=0.0614 and a graph as shown in Figure 33.

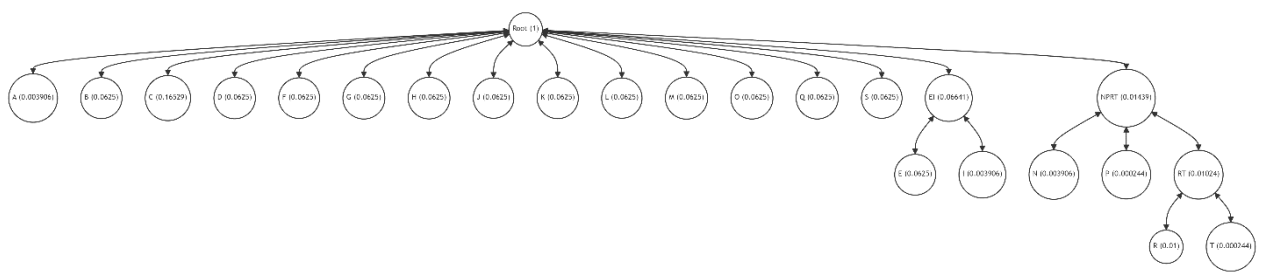

**Figure 31:** Fragment of a Patricia-Merkle Tree with $m=16$

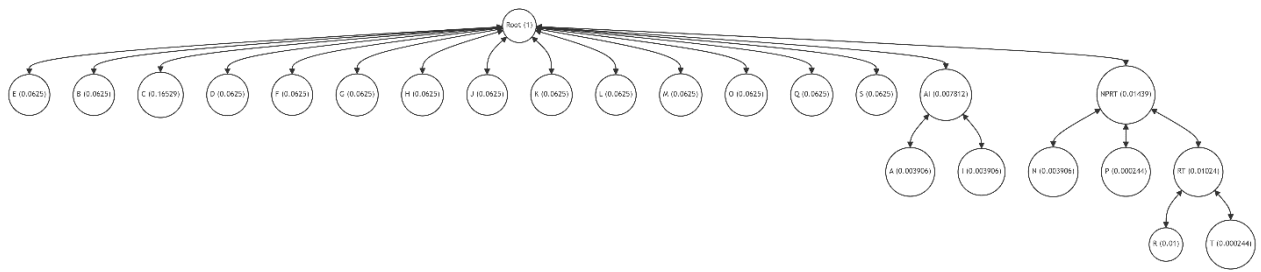

**Figure 32:** Result of the First Optimization of the Patricia-Merkle Tree with $m=16$

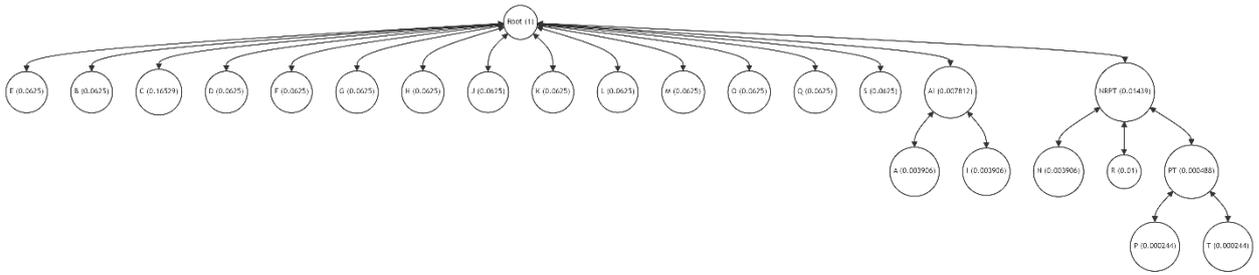

**Figure 33:** Result of the Second Optimization of the Patricia-Merkle Tree with $m=16$

Thus, even a small number of algorithm iterations allows for a significant reduction in discrepancy (1) and optimization of the tree structure, reducing the average path length.

This example underscores the potential of our restructuring algorithm to enhance the efficiency of Patricia-Merkle trees in blockchain applications. By judiciously swapping the positions of leaf pairs, we can significantly improve the tree's structure, aligning it closer to the optimal configuration. This process not only minimizes the average path length but also contributes to the overall efficiency and scalability of blockchain operations, particularly in systems like Ethereum where Patricia-Merkle trees play a crucial role in data integrity verification.

**7. Path Encoding in the Adaptive Merkle Tree**

The integration of adaptive Merkle Trees into existing blockchain systems like Ethereum presents a paradigm shift in data integrity verification. This shift, while promising significant efficiency gains, also necessitates substantial modifications to current protocols. In Ethereum's existing structure, an account's address directly determines its encoding path in the Patricia-Merkle Tree. This encoding, defined by a series of nibbles (four-bit blocks), uniquely maps each address from the root to a specific leaf in the vast tree structure. The current system's design allows for the seamless integration of new addresses into this expansive tree.

Adopting an adaptive approach fundamentally alters this scenario. Instead of a balanced structure, we would deal with a highly unbalanced tree where frequently used leaves are positioned closer to the root, and less probable leaves are relegated to lower levels. Implementing this directly in the existing Patricia-Merkle Tree structure is not feasible. However, creating a new tree during a protocol update in Ethereum could allow for the incorporation of this adaptive approach, radically changing the concept of path encoding in the tree.

The challenge arises in reconciling existing addresses with new path encodings. In the new structure, a random address would no longer be tied to a specific path encoding but would merely determine the leaf's value, not its path from the root. This concept is illustrated in Figures 34 and 35, where Figure 34 shows the simplified path encoding in a balanced tree with corresponding addresses, and Figure 35 depicts these addresses in an adaptive tree with Huffman code-based path encodings.

A practical solution to address compatibility issues in the adaptive tree is the tabular storage of two structures: "account address – path encoding." This approach allows for the adaptation of the tree based on the usage probabilities of addresses, leading to significant savings in verification complexity and cost. Simultaneously, it preserves the existing mechanism for generating random addresses, including the ability to transfer funds to not-yet-created accounts.

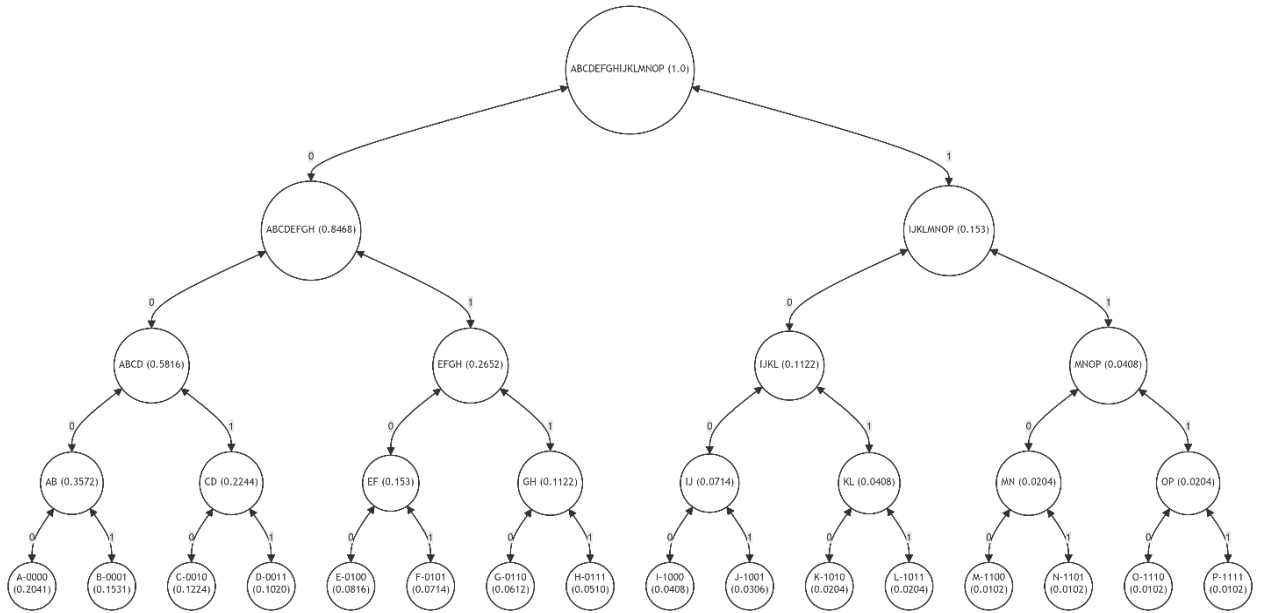

**Figure 34:** Path Encoding in a Balanced Merkle Tree

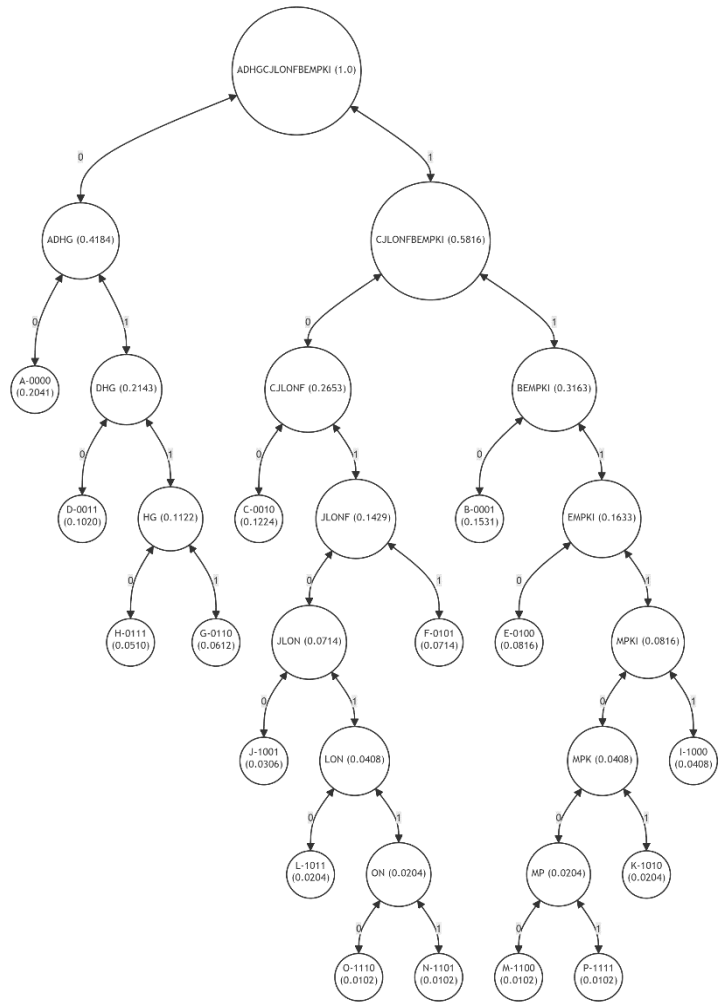

**Figure 35:** Adaptive Merkle Tree with Huffman Code-Based Path Encoding

Figure 34 illustrates the path encoding mechanism within a traditional balanced Merkle Tree, as utilized in current blockchain systems like Ethereum. Each account address is directly linked to a unique path encoded by a series of nibbles, efficiently mapping the journey from the tree's root to the respective leaf. This representation underscores the systematic and predictable nature of path encoding in a balanced tree structure, highlighting the ease with which new addresses can be integrated into the expansive tree.

Figure 35 depicts the transformative approach of an adaptive Merkle Tree, where path encodings are based on Huffman codes. This figure contrasts sharply with Figure 34, showcasing a more dynamic and usage-frequency-oriented structure. In this adaptive model, the path encoding is no longer a straightforward derivative of the account address but is instead determined by the frequency of data access, leading to a highly unbalanced but efficient tree structure. This figure effectively demonstrates the shift from a uniform to a tailored approach in path encoding, aligning more closely with the actual usage patterns within the blockchain network.

**Table 6:** Correlation between Account Addresses and Path Encodings in Adaptive Merkle Tree

| Leaves | Leaf probabilities | Path Encoding in a Balanced Merkle Tree (Account Addresses) | Huffman Code-Based Path Encoding in Adaptive Merkle Tree |
|---|---|---|---|
| A | 0.2041 | 0000 | 00 |
| B | 0.1531 | 0001 | 110 |
| C | 0.1224 | 0010 | 100 |
| D | 0.1020 | 0011 | 010 |
| E | 0.0816 | 0100 | 1110 |
| F | 0.0714 | 0101 | 1011 |
| G | 0.0612 | 0110 | 0111 |
| H | 0.0510 | 0111 | 0110 |
| I | 0.0408 | 1000 | 11111 |
| J | 0.0306 | 1001 | 10100 |
| K | 0.0204 | 1010 | 111101 |
| L | 0.0204 | 1011 | 101010 |
| M | 0.0102 | 1100 | 1111000 |
| N | 0.0102 | 1101 | 1010111 |
| O | 0.0102 | 1110 | 1010110 |
| P | 0.0102 | 1111 | 1111001 |
| Average code length | | 4 | **3.49 (13% more efficient)** |

Table 6 presents a crucial solution to the challenge posed by the transition to an adaptive Merkle Tree: a tabular representation of the correlation between account addresses and their new path encodings. This table exemplifies the practical approach to reconciling the existing system of address generation with the innovative path encoding mechanism of the adaptive tree. By maintaining a record of these correlations, the table ensures that the integrity and functionality of the blockchain are preserved, even as the system evolves to embrace more efficient data verification methods. This tabular approach symbolizes a bridge between the legacy structures of blockchain and the future-oriented adaptive Merkle Tree, ensuring a seamless transition and operational continuity.

## 8. Enhancing Verkle Trees Through Adaptive Restructuring

The advent of Verkle trees represents a significant leap forward in the optimization of blockchain storage and verification processes. By combining the succinctness of vector commitments with the hierarchical structure of Merkle trees, Verkle trees offer a promising solution to scalability and efficiency challenges in blockchain systems. This section delves into the potential applications of our adaptive restructuring approach to Verkle trees, exploring how dynamic adjustments to tree configurations can further enhance their efficiency and applicability in blockchain technologies.

## 8.1. Application of Adaptive Trees in Verkle Tree Technology

Verkle trees, a novel data structure, merge the benefits of Merkle trees with vector commitments, providing a compact, efficient means of storing and verifying blockchain state. They stand poised to revolutionize data storage in blockchain by significantly reducing the size of proofs required for state verification. Our approach, centered on adaptive restructuring, introduces a method to dynamically adjust Verkle tree configurations based on usage patterns, thereby optimizing both storage efficiency and verification speed.

Adaptive restructuring in the context of Verkle trees involves the dynamic adjustment of tree branches and nodes based on the frequency and patterns of data access and updates. This method leverages statistical analysis to predict which parts of the tree are accessed more frequently, allowing for a more efficient organization of data. By applying Huffman or Shannon-Fano coding principles, we can ensure that the most accessed elements are closer to the root, thereby reducing the path length for common operations.

## 8.2. Technology and Advantages

- Reduced Proof Sizes: By optimizing the structure of Verkle trees to reflect access patterns, we can significantly reduce the size of proofs required for verifying transactions. This is because frequently accessed data can be positioned closer to the root, making it quicker and less resource-intensive to generate and verify proofs.

- Enhanced Verification Speed: Adaptive restructuring can lead to a more efficient verification process. Shorter paths for frequently accessed data mean that less computational effort is required to verify transactions, enhancing the overall throughput of the blockchain network.

- Dynamic Scalability: As blockchain systems evolve, so do their storage and access patterns. Adaptive restructuring allows Verkle trees to dynamically adjust to these changes, ensuring that the data structure remains optimized for current usage trends. This adaptability is crucial for maintaining high performance as the system scales.

- Cost Efficiency: By optimizing the path lengths for data access and verification, the proposed approach can also reduce the cost associated with these operations. In blockchain systems where transaction costs are a significant concern, such as Ethereum, this can lead to substantial savings for users and applications.

- Application in Sharding: Verkle trees are particularly well-suited for sharded blockchain architectures. Adaptive restructuring can enhance the efficiency of cross-shard communication by optimizing the storage and retrieval of shard-specific data, further improving the scalability of sharded networks.

Thus, the integration of adaptive restructuring techniques with Verkle tree technology presents a promising avenue for enhancing blockchain efficiency. By dynamically optimizing data storage and access patterns, we can achieve significant improvements in proof size, verification speed, and overall system scalability. This approach not only addresses current scalability and efficiency challenges but also provides a flexible framework that can adapt to future developments in blockchain technology. As we continue to explore the potential of adaptive Verkle trees, it becomes increasingly clear that this innovative approach could play a pivotal role in the next generation of blockchain systems.

## 9. Discussion

In this work, we have embarked on a comprehensive exploration of optimizing tree structures within the blockchain ecosystem, addressing the critical challenge of scalability that plagues current blockchain technologies. Our investigation spans from conceptualizing the inherent problems associated with traditional Merkle trees to proposing and validating an innovative approach for adaptive restructuring of these trees to enhance efficiency and scalability in blockchain systems.

The blockchain paradigm, while revolutionary, faces significant scalability challenges, primarily due to the inherent limitations of its underlying data structures and consensus mechanisms. Traditional Merkle trees, despite their widespread adoption for ensuring data integrity and facilitating efficient verifications, contribute to these scalability issues due to their static nature and the increasing cost of operations as the blockchain grows.

Existing solutions to blockchain scalability, such as sharding and layer 2 protocols, offer partial remedies by distributing the workload or offloading transactions. However, these approaches often introduce complexity or compromise on decentralization and security. Our review of the state of the art highlights a gap in dynamically optimizing the data structures themselves to directly address the root causes of inefficiency.

### 9.1. Our Contribution

Our primary contribution lies in the introduction of adaptive Merkle trees, a novel concept that leverages dynamic restructuring based on usage patterns to optimize path lengths and reduce the computational overhead associated with data verification and integrity checks. By applying principles from Huffman and Shannon-Fano coding to the organization of tree nodes, we ensure that frequently accessed data is more accessible, thereby reducing the average path length and associated costs.

Through rigorous analysis and examples, we demonstrated the efficiency gains achievable with adaptive Merkle trees. Our algorithm for Merkle tree restructuring, detailed in Section 5, provides a systematic approach for dynamically adjusting tree structures, significantly improving upon the static nature of traditional Merkle trees.

Extending our concept to Verkle trees, we showcased how adaptive restructuring could be applied to this advanced data structure, further enhancing its efficiency and making it even more suitable for large-scale blockchain applications. This application not only underscores the versatility of our approach but also its potential to contribute to the next generation of blockchain technologies.

### 9.2. Comparison with Existing Solutions

In the quest to address blockchain scalability, several innovative solutions have been proposed and implemented across various platforms. Each of these solutions presents unique advantages and challenges. Below (Table 7), we provide a comparative analysis of these solutions, including our adaptive restructuring approach, to highlight their relative strengths and limitations.

**Table 7:** Comparison of Scalability Solutions in Blockchain Technology

| Solution Type | Examples | Advantages | Disadvantages |
|---|---|---|---|

| Sharding | Ethereum 2.0, Zilliqa | - Distributes workload across multiple chains.<br>- Enhances transaction throughput. | - Increases complexity.<br>- Potential security risks due to smaller validator sets. |
|---|---|---|---|
| Layer 2 Protocols | Lightning Network, Plasma | - Offloads transactions from the main blockchain.<br>- Facilitates faster and cheaper transactions. | - Can introduce centralization points.<br>- Complex to manage and integrate. |
| State Channels | Raiden Network, Celer Network | - Enables off-chain transaction channels.<br>- Instantaneous transaction settlement. | - Requires on-chain settlement for disputes.<br>- Limited to participants in the channel. |
| Sidechains | Liquid Network, POA Network | - Allows for customizable blockchains linked to the main chain.<br>- Facilitates specific use cases and scalability. | - Security is often reliant on the main chain.<br>- Interoperability challenges. |
| **Adaptive Merkle Trees** | **Our Approach** | **- Dynamically optimizes data structure based on usage.**<br>**- Reduces average path length and verification costs.** | **- Requires initial restructuring and maintenance.**<br>**- Concept is newer and less tested in real-world scenarios.** |

The comparative analysis underscores the diversity of approaches to tackling blockchain scalability, each with its unique trade-offs. Sharding and Layer 2 protocols, while promising significant throughput improvements, introduce additional layers of complexity and potential security concerns. State channels and sidechains offer more specialized solutions but are limited by their applicability and integration challenges.

Our approach, adaptive restructuring of Merkle and Verkle trees, stands out by directly optimizing the underlying data structure of the blockchain. This method offers a fundamental improvement in efficiency without introducing external dependencies or significantly altering the blockchain's operational principles. While it necessitates initial efforts for restructuring and ongoing maintenance, the benefits of reduced path lengths and lower verification costs present a compelling case for its adoption. Moreover, being a relatively new concept, it opens up extensive opportunities for further research and development to fully realize its potential and address any emerging challenges.

Thus, our work contributes a novel perspective to the field of blockchain research, opening new avenues for the development of more scalable and efficient blockchain systems. By addressing scalability at the data structure level, we provide a foundational solution that can be integrated with other scalability and efficiency-enhancing techniques, offering a comprehensive approach to overcoming one of the most significant barriers to blockchain adoption.

## 10. Conclusion

The exploration of adaptive restructuring in Merkle and Verkle trees within this study presents a novel approach to addressing the enduring challenge of blockchain scalability. By dynamically adjusting the structure of these trees based on usage patterns, we propose a method that significantly reduces the average path length for verification processes, thereby enhancing the efficiency and scalability of blockchain systems.

Our contribution to the field of blockchain technology is twofold. Firstly, we introduce a conceptual framework for the adaptive restructuring of Merkle trees, which lays the groundwork for practical implementations in existing blockchain infrastructures. Secondly, through a series of detailed examples, we demonstrate the feasibility and benefits of our approach, highlighting its potential to optimize verification processes and reduce associated costs.

Comparative analysis with existing scalability solutions reveals that while many approaches offer improvements in transaction throughput and efficiency, they often introduce additional complexity or security concerns. In contrast, adaptive restructuring directly targets the underlying data structure of the blockchain, offering foundational improvements without compromising on security or introducing external dependencies.

The implications of our research extend beyond theoretical advancements. By providing a scalable and efficient method for data verification, adaptive restructuring has the potential to facilitate broader adoption of blockchain technology across various sectors, including finance, supply chain management, and beyond. It opens up new avenues for blockchain applications that require high throughput and efficient data integrity verification.

In conclusion, the adaptive restructuring of Merkle and Verkle trees represents a significant step forward in the quest for blockchain scalability. It offers a unique blend of efficiency, security, and practicality, making it a promising solution for the next generation of blockchain systems. As the blockchain ecosystem continues to evolve, the principles and methodologies outlined in this study will undoubtedly contribute to its growth and maturity, paving the way for more scalable, efficient, and versatile blockchain architectures.